\documentclass[manuscript=perspective,layout=twocolumn,email=false,maxnames=50,arial,10pt]{achemso}

\makeatletter
\let\l@addto@macro\relax
\makeatother
\usepackage[fontsize=10pt]{scrextend}
\setkeys{acs}{articletitle = false}
\usepackage{graphicx}
\usepackage{mathtools}
\usepackage{graphicx}   
\usepackage{color}
\usepackage{ulem}
\usepackage{enumitem}
\setlist{leftmargin=2mm}
\usepackage[caption=false]{subfig}
\usepackage{lineno}
\usepackage{setspace}
 \SectionNumbersOn

\usepackage{lipsum}
\usepackage{etoolbox}

\usepackage{array}
\usepackage{booktabs}
\usepackage{multirow}

\usepackage{booktabs}

\makeatletter
\AtBeginDocument{\preto\@title{\singlespacing}}
\makeatother

\title{Ligand dissociation mechanisms from all-atom simulations: Are we there yet?}
 \author{Jo\~{a}o Marcelo Lamim Ribeiro}
 \affiliation{Department of Chemistry and Biochemistry and Institute for Physical Science and Technology,
 University of Maryland, College Park 20742, USA.}
\author{Sun-Ting Tsai}
 \affiliation{Department of Physics and Institute for Physical Science and Technology,
 University of Maryland, College Park 20742, USA.}
  \author{Debabrata Pramanik}
 \affiliation{Department of Chemistry and Biochemistry and Institute for Physical Science and Technology,
 University of Maryland, College Park 20742, USA.}
 \author{Yihang Wang}
 \affiliation{Biophysics Program and Institute for Physical Science and Technology,
 University of Maryland, College Park 20742, USA.}
 \author{Pratyush Tiwary}
 \affiliation{Department of Chemistry and Biochemistry and Institute for Physical Science and Technology,
 University of Maryland, College Park 20742, USA.}
 \email{ptiwary@umd.edu} 

\let\oldmaketitle\maketitle
\let\maketitle\relax

\begin{document}
\twocolumn[
  \begin{@twocolumnfalse}
    \oldmaketitle
    \begin{abstract}
Large parallel gains in the development of both computational resources as well as sampling methods have now made it possible to simulate dissociation events in ligand-protein complexes with all--atom resolution. Such encouraging progress, together with the inherent spatiotemporal resolution associated with molecular simulations, has left their use for investigating dissociation processes brimming with potential, both in rational drug design, where it can be an invaluable tool for determining the mechanistic driving forces behind dissociation rate constants, as well as in force-field development, where it can provide a catalog of transient molecular structures on which to refine force-fields. Although much progress has been made in making force-fields more accurate, reducing their error for transient structures along a transition path could yet prove to be a critical development helping to make kinetic predictions much more accurate. In what follows we will provide a state-of-the-art compilation of the molecular dynamics (MD) methods used to investigate the kinetics and mechanisms of ligand-protein dissociation processes. Due to the timescales of such processes being slower than what is accessible using straightforward MD simulations, several ingenious schemes are being devised at a rapid rate to overcome this obstacle. Here we provide an up-to-date compendium of such methods and their achievements/shortcomings in extracting mechanistic insight into ligand-protein dissociation. We conclude with a critical and provocative appraisal attempting to answer the title of this review.
    \end{abstract}
  \end{@twocolumnfalse}
]

\maketitle
\section{INTRODUCTION}
Recent decades have seen enormous gains in the development of computational hardware both in general-purpose computing resources such as graphical processing units (GPUs) and in molecular simulation-specific hardware. \cite{dror2012biomolecular,shaw2010atomic} Together these have pushed the timescales that are now accessible into a regime where important biochemical processes can be studied using all-atom molecular dynamics (MD) simulations. As a result a widening catalog of biochemical processes have been studied at the atomistic level,\cite{dror2012biomolecular,faraldo2007importance,berteotti2008protein,dror2009identification,vanni2009observation,lyman2009role,kresten_elife,seibert2005reproducible,simmerling2002all,voelz2010molecular,shaw2010atomic,arkin2007mechanism,bostick2007selectivity,enkavi2010simulation,jensen2010principles,noskov2004control} including the dissociation of ligand-macromolecule complexes.\cite{dror2011gpcr,shaw_dasatinib,hurst2010lipid,mondal2018atomic} These dissociation processes are relevant in various contexts, including developing a fundamental understanding for the chemical basis of life processes as well as in the rational design of drugs. \cite{dror2012biomolecular,zeevaart2008optimization} With respect to the latter, for instance, MD simulations have been found to be ever more accurate in predicting the relative binding free energies $\Delta \Delta G_{b}$ between families of ligands and/or receptors (see Table \ref{tab:table-defi} for definitions of $\Delta \Delta G_{b}$ and other terms).\cite{deng2006calculation,deng2009computations,wang2015accurate}

In particular, the determination of accurate mechanisms for ligand-protein dissociation processes, which can happen over a timespan of microseconds to several hours,\cite{copeland2006drug,copeland2016drug,pan_kinetics,fullerene,sgoop_fullerene,wang2017biomolecular} is a classic unresolved biochemical problem. There are experimental techniques that have been developed capable of measuring the overall rate constants for ligand-protein dissociation processes.\cite{copeland2006drug} It remains extremely difficult, however, for such experiments to glean an atomic residue-by-residue level 
understanding of the dissociative mechanism. Much of the complications -- and richness -- of the problem arise from the inherent ever-fluctuating structures, including those of the protein, ligand and the solvent medium. Furthermore, recently it has been highlighted\cite{tummino2008residence,trypsin,copeland2006drug,copeland2016drug} that a very important aspect of protein-ligand interactions is the ligand's residence time in the target protein --  in many systems the residence time, which correlates more with mechanism, is a much strong predictor of eventual function than the thermodynamic affinity, which correlates more with structure. Often the residence time is quantified through its reciprocal, the dissociation rate constant $k_{off}$ (see Table \ref{tab:table-defi}). MD simulations could, in principle, be ideal vehicles for investigating these dissociation process with all-atom resolution, but their relevant timescales lie far past what can be simulated even with the best computing resources if one was to use straightforward MD simulations. This is because complex biochemical processes often involve multiple metastable states in the presence of high barriers, while the fast vibrational motions constrain the integration time step in a MD simulation to femtosecond (fs) values.

The idea behind this review is to perform a critical examination of where we stand in the context of using all-atom MD simulations for studying the dissociation process of protein--ligand complexes. This includes the calculation of their association/dissociation rate constants, the metastable and transition states relevant to the process, and various other related biochemical factors. In order to tackle problems plagued with such extreme rare events, several specialized methods have been proposed and used to deal with the long timescales required to observe the desired rare event in a simulation.\cite{umbrella,meta_laio,rave}  MD simulations thus act as \textit{in silico} microscopes revealing the mechanisms giving rise to the macroscopic rate constant $k_{off}$ still often determined from experiments. It can provide a full catalog of intermediate and transition states that are hard to characterize in experiments due to their transient nature, revealing specific molecular attractions, steric hindrances, macromolecular motions and water/solvent effects that together give rise to the measured rate constant. This attribute has the potential of making it an invaluable tool in rational drug design.

Here we provide a state--of--the--art glossary and overview of many of the methods that have been proposed for the purpose, highlighting their key underlying principles, and what we believe to be their key strengths and weaknesses. In parallel, we provide specific case--studies demonstrating the use of MD methods for studying the kinetics of ligand dissociation. We finally conclude with our summary of the field -- which is half parts optimistic, and half parts concerned that there are many challenges still to be surmounted. We believe this review should serve as the definitive manual for someone willing to develop or apply such methods, as well as for someone just trying to form an opinion on which method to use for a specific calculation. 

\begin{table}[H]
\vspace{-5pt}
\begin{center}
\begin{tabular}{|>{\raggedright}p{.6in}
                |p{2.5in}|}  
\hline
{\bf Term}  & {\bf Definition} \\ 
\hline
$\Delta G_{b}$ & Absolute binding free energy \\
\hline
 $\Delta \Delta G_{b}$  & Relative binding free energy difference between two systems\\
\hline
$k_{off}$ & Ligand-macromolecule complex dissociation rate constant \\ 
\hline
$k_{on}$ & association rate constant equaling $e^{-\beta \Delta G_{b}} k_{off} $ \\ 
\hline
 \textbf{x} & Molecular configuration of ligand-macromolecule complex \\ 
\hline
 RC & Reaction coordinate \\ 
\hline
\end{tabular}
\end{center}
 \caption{\label{tab:table-defi}Definition of important terms.}
\end{table}
\renewcommand{\baselinestretch}{1.0}

\section{METHODS}
\subsection{Unbiased methods}
As long as a sufficient number of unbinding events are observed in an unbiased MD simulation, the dissociation rate constants and mechanisms are straightforward to obtain through simple counting. Of course the timescale problem limits this approach to just ligand-protein complexes that are fast (un)binders, although the introduction of the Anton supercomputer\cite{shaw2008anton,shaw2009anton}, which was designed for and can run MD simulations two orders of magnitude faster than conventional supercomputers, has 
increased the number of association processes that can be studied via long unbiased simulations\cite{shaw_dasatanib,dror2012annrevbios,larsen2012jacs,pan_kinetics}. It remains difficult, however, to use long unbiased MD for ligand dissociation processes due to their much slower rates in general.\cite{copeland2006drug,copeland2016drug,pan_kinetics} Dissociations studied via this brute-force approach are thus constrained to just the fastest unbinders. For example, \citet{caflisch_plos} have investigated the dissociation of small ligand molecules from the FKBP protein binding sites through unbiased MD simulations in explicit water. \citet{pan2017jctc} ran long simulation on Anton for the unbinding of several small ligands from the FKBP protein. In this section, we summarize a number of methods that do not require the explicit biasing of an MD simulation. These methods, in one way or another, are designed to patch several unbiased simulations to form a picture of the dissociation process. In the authors' view, it is non-trivial in most of these methods to truly assess if especially for slow unbinders, such a process can lead to truly unbiased results (see Sec. \ref{discussion}).

\subsubsection{Markov State Models}
Markov State Models (MSMs) provide a powerful framework with which to model the kinetics of biochemical processes. One of its appealing features is that it models the kinetics through much shorter MD simulations run in parallel,\cite{prinz2011markov,chodera2014markov,pande2010everything} which in principle helps to ameliorate the difficulties with calculating kinetic estimates of rare events with statistical confidence. In essence, an MSM is a parametric model for describing the long-time dynamics of a complex system in terms of transitions between discrete states in configuration space.\cite{pande2010everything} The parametrization of an MSM then involves two crucial steps: (a) the definition of the states relevant to describing the dynamics of interest and (b) calculating the transition probabilities for interconverting between these states. An important point to keep in mind is that MSMs often define thousands of microstates (using clustering algorithms on MD simulation data) along some small collection of collective variables. It is then possible to ``stitch'' together these short trajectories to estimate the transition probability matrix between the states as long as the system is evolving under a Markovian process, where the transitions into future states will depend just on the system's current state, regardless of its past history. This will be true if one is willing to coarsen the time-resolution of viewing the system, also called lag time. In practice, approximate Markovianity is reached if a long enough lag time is chosen, and hence it is an important part of building a MSM.

Once the transition probability matrix is determined, its eigenvalues and eigenvectors also provide us information of the dynamical properties of the system. Convergence of the eigenvalues with choice of lag time $\tau$ has been the traditional test for ascertaining Markovianity. There have been several efforts made in improving the discretization of configuration space into distinct states, which is the most important and nontrivial step in building a MSM. For years, people have known that the best set of states would be the one with the highest barriers separating states. Choosing the best Markov states is then equivalent to finding out the slowest processes on the free energy landscape. This criterion guided MSM building in many studies. Recently, N\`{o}e and N\"{u}ske introduced a variational analysis for the approximation of eigenvalues and eigenfunctions of the transition matrix, called the variational approach to conformational dynamics. Another approach using time structure-based (time-lagged) independent component analysis (tICA) has also been proposed to construct MSMs. Later on, P\'{e}rez-Hern\'{a}ndez et. al. have shown that tICA can be used to provide an optimal solution to the variational approach to conformational dynamics\cite{noe_jcp_2013}, and hence can be used to determine the states of MSMs. A crucial open problem in this sub-field is the use of biased simulations for constructing MSMs, an area of great active interest.\cite{mey2014xtram,stelzl2018dynamic}

\subsubsection{Weighted ensemble}
Weighted ensemble (WE) uses the idea of resampling to speed-up the observation of a desired rare event.\cite{bhatt2010steady,zhang2010weighted,zuckerman2017weighted,huber1996weighted,dickson2016ligand,dickson2017multiple,nunes2018escape} In order to initiate a WE calculation, the first step is to launch an ensemble of trajectories drawn from some initial distribution in configuration space, $P=P(\textbf{x})$, which might be time-dependent or independent.\cite{zuckerman2017weighted} Once a short amount of time has passed the trajectories are re-evaluated for their progress along the binned configuration space, with certain configurations copied and others eliminated based on a rigorous adjustment of their weights made to ensure this new sample of configurations will conserve the current value of $P(\textbf{x})$.\cite{zuckerman2017weighted} This resampling in WE thus aims to increase the number of trajectories visiting the regions of configuration space with low probabilities. The WE method will then re-initiate the ensemble of trajectories from the current configurations and proceed to iterate through the resampling and MD simulation steps until the desired rare event is observed.\cite{zuckerman2017weighted}

One important point to consider is that the binning of configuration space in WE simulations will be specific to the problem at hand.\cite{zuckerman2017weighted} This predefinition of bins might in turn affect the ideal values of other parameters such as the time lag between resampling steps.\cite{zuckerman2017weighted} It is also worth mentioning, however, that the bins do not have to be fixed during the a WE simulation.\cite{zhang2010weighted,huber1996weighted,zuckerman2017weighted} For instance, there exist adaptive WE simulation schemes allowing the center and the size of bins to change\cite{zhang2010weighted}. The bins learn to maximize the distance between each other within the space that has been sampled and will spread out as more space has been sampled.\cite{zhang2010weighted}

It is also important to mention that the parallelization of trajectories in the WE method make extracting macroscopic rate constants not so straightforward.\cite{bhatt2010steady,zuckerman2017weighted} This is due to the fact that parallelization decreases the relaxation time available to each simulation in order to help the system reach steady-state.\cite{zuckerman2017weighted} \citet{bhatt2010steady} have devised an approach from which steady-state kinetic information such as the mean first-passage time can be calculated however.

\subsubsection{Milestoning}
The idea behind the milestoning method is to divide configuration space into several cells\cite{vanden2009markovian,bello2015exact} in order to break down the dynamics of the system into the local transitions between adjacent cells.\cite{vanden2008assumptions} In its original implementation, the division of configuration space was performed along an approximate reaction coordinate (RC), with each discretized point along this RC then defining a $3N - 1$ dimensional hypersurface called a milestone.\cite{milestoning} In more recent formulations a milestone is defined as a dividing surface between two cells.\cite{bello2015exact} The transitions between milestones represent the most critical event in this class of methods and short MD simulations in parallel are initiated at a given milestone and stopped when the simulation reaches an adjacent milestone.\cite{milestoning,vanden2008assumptions,bello2015exact} These local transition events can be ``stitched'' together to recover the long-time kinetic properties of the rare event process.\cite{vanden2008assumptions,bello2015exact} It is worthwhile to mention that the choice of milestones have often satisfied the condition that the transitions between them are Markovian,\cite{vanden2008assumptions,vanden2009markovian} the consequence of being that the first-passage-time distribution will be independent of the full history before arriving at a given milestone. A recent development is the introduction of a formally exact formulation where the mean first passage time was derived from exact statistical analysis of the individual local mean first passage times for the functional transitions between the conformational space cells.\cite{bello2015exact}

\subsubsection{Adaptive multilevel splitting}
\label{ams}
In adaptive multilevel splitting (AMS)\cite{trypsin_schulten}, the generic objective is to calculate the mean first passage time corresponding to a transition from an initial reactant metastable state, $A$, to a final product metastable state, $B$. For this AMS aims to estimate the committor probability, $P_{c}$, known to be the exact RC for any arbitrary reaction, and roughly defined as probability that a trajectory initiated in the reactant state will reach the product state before returning.\cite{peters_rc} With the estimated $P_{c}$, the mean first passage time corresponding to a transition from $A$ to $B$ can be calculated. Although $P_{c}$ is the perfect RC, it is important to keep in mind that it is not known a priori -- it is after all the quantity that the AMS algorithm is attempting to estimate. The AMS algorithm requires that progress along a RC be measured and an approximate user-defined trial RC  $z$ different than $P_{c}$ is used for this. The basic AMS algorithm begins with $N$ initial MD trajectories launched from a point $z = z_{0}$, denoting an initial value corresponding to the reactant state $A$. The MD simulations are all stopped when the $N$ replicas return to a RC value $z = z_{0}$. The next step in the AMS algorithm is to examine the progress that each replica made in sampling the RC during the simulation. For this, AMS involves systematic removal and addition of replica, until the desired rare event from A to B has been sampled and an estimate of $P_{c}$ from the product of different conditional probabilities for moving between different $z$ values can be calculated. $P_{c}$ is then directly involved in calculating the rate constant through a simple expression.\cite{trypsin_schulten}

\subsection{Biased sampling based methods}
Due to the timescale associated with ligand unbinding processes often being minutes or longer,\cite{copeland2006drug,dickson2017kinetics} a large number of studies have sought to  introduce biasing forces into the molecular description of the system to drive the simulation to sample the process of interest.\cite{umbrella,meta_laio,wtm,rave} Care must be taken, however, to do so in a rigorous manner allowing the statistics of the original, unperturbed process to be recovered.\cite{tiwary_rewt} This is relatively straightforward to do when the aim is to recover static equilibrium properties such as free energies, but it turns out to be a much more formidable task when the goal is to recover, from a biased simulation, the unbiased kinetics, although methods to do so have been developed.\cite{meta_time} Unfortunately, the need to introduce bias into molecular simulations can often be severe when one is interested in capturing accurate rate constants and associated mechanisms. The reason is that several rare events must be observed in order to obtain kinetic estimates that are of statistical significance.\cite{bowman2016accurately} While binding free energies are state functions that require appropriate sampling of just the initial and final states\cite{deng2006calculation,deng2009computations} of the desired biochemical process, the kinetics are path-dependent, meaning a whole ensemble of transition paths\cite{throwingropes} must be sampled before an averaged kinetic estimate can be produced. Here we summarize one such very popular method, namely metadynamics, that makes it possible to get unbiased rate constants and pathways from biased MD simulations. Note that there are alternative approaches to adding a biasing potential by increasing the temperature of the system, either whole or along parts, for which we refer the reader to another recent thorough review of methods.\cite{dickson2017kinetics}
  
The metadynamics\cite{arpc_meta} method has been the most popular of these biasing approaches due to the development of the infrequent metadynamics framework, which allows unbiased kinetics to be extracted from the biased simulation using an acceleration factor\cite{meta_time}. The basic idea behind metadynamics is to add a history-dependent bias potential to the system's Hamiltonian using a predetermined deposition stride, and as a function of a set of collective variables (CVs) $s(\textbf{x})$ mapping the high-dimensional $\textbf{x}$ into a low-dimensional representation. The bias $V(s,t)$ is typically constructed as a sum of gaussians deposited along the trajectory in the CV space. As time goes by, this additional bias prevents the system from revisiting the same site in the CV space. The free energy surface (FES) $F(s)$ is then obtained from the negative of the bias potential added. If the FES is truly well-converged, especially in the barrier regions, then the kinetic rate constants can also be estimated from the recovered barrier height between the metastable states by transition state theory (TST). However, a fundamental issue in using a FES derived from metadynamics, umbrella sampling or any other method for rate calculation is that of dynamical corrections to the TST rate.\cite{tiwary2016review} The transition rate calculated by TST is only a crude upper bound to the true rate, and a transmission coefficient needs to be calculated for example by launching trajectories from top of every barrier identified on the FES. 

A much simpler method named ``infrequent metadynamics" was proposed to get unbiased dynamics from biased dynamics which bends around the calculation of the transmission coefficient by making the assumption that it is defined only by the properties of the barrier. \cite{meta_time} Thus if one could avoid adding bias on the barrier, the transmission coefficient could be kept same between the biased and unbiased simulations. Infrequent metadynamics allows doing this assuming a decent CV is known for the process being stand. Note that this CV does not have to be the perfect RC (see Sec. \ref{ams} for descritpion of committor probability) but can be a simple combination of order parameters which can be optimized using preliminary $frequent$ metadynamics runs with the method SGOOP \cite{sgoop,tiwary_biotin} Once such a CV is known, infrequent metadynamics proceeds by simply decreasing the bias addition frequency so that the time interval between two bias addition events becomes slower than the time spent in the barrier regions, thereby avoiding adding bias there. The unbiased timescale can then be obtained from the biased timescale simply through the calculation of an acceleration factor, which is the running average of the exponentiated bias given by $\alpha(t)=(1/t)\int^{t}_{0}e^{\beta V(s)}dt$ where $s$ is the CV being biased and $t$ is the simulation time. The reliability of the unbiased dynamics so reconstructed can then be ascertained through a simple p-value test.\cite{pvalue} The idea of using acceleration factor in metadynamics to directly obtain unbiased kinetic information has also been introduced in other variants which we do not discuss here. \cite{mccarty2015variationally,fu2017determining,bortolato2015gpcr,wang2018frequency}

\subsection{Machine learning based methods}
Recently, attempts have been made to use machine learning to ameliorate the timescale problem in all-atom simulation of biochemical processes.\cite{noe_timelaggedautoenc,sultan2018transferable,hernandez2018variational,rave,ribeiro2018achieving} Although these machine learning protocols have been leveraged in a few different contexts, the most common approach involves using neural networks trained to learn a low-dimensional representation approximating the RC.\cite{noe_timelaggedautoenc,noe_vampnet,sultan2018transferable,hernandez2018variational} Finding such low-dimensional RC representations is often a crucial step in both the biased and unbiased algorithms we have described. Two particular approaches have directly employed the neural network based RC for use in an MSM,\cite{noe_timelaggedautoenc,noe_vampnet,hernandez2018variational} or for use with a biased MD method like metadynamics\cite{sultan2018transferable} or umbrella sampling.\cite{chen2018collective}
For instance, \citet{noe_vampnet} suggested the use of two parallel neural networks in conjunction with the variational principle for Markov process (VAMP) to propose VAMPnets,\cite{noe_vampnet} which in principle can automate the generation of an MSM. Building an MSM often requires technical expertise on the part of the investigator and could be an important development in increasing the reach of MSM. A different approach,\cite{rave,ribeiro2018achieving} which our group has proposed, uses an unsupervised machine learning method to learn both a low-dimensional approximate RC as well as a bias potential along this neural network RC, and hence is its own enhanced sampling method. One interesting feature of this approach, named Reweighted autoencoded variational Bayes for enhanced sampling (RAVE),\cite{rave,ribeiro2018achieving} is that the RC and bias potential are learnt from the neural network \textit{simultaneously}, which turns out to be helpful in screening through spurious machine learning solutions of the biasing parameters. The RAVE protocol involves iterating between machine learning and MD, with each iteration learning a more refined low-dimensional RC representation and bias potential, these iterations continuing until these biasing parameters are converged and sufficient to sample the desired ligand dissociation. One benefit of this approach is that a time-independent bias is produced which can be used to launch multiple independent production MD runs, which is helpful in extracting statistically robust kinetics information. Thus far, RAVE has been applied to the benzene-lysozyme complex in order to calculate absolute binding affinities in close agreement with other enhanced sampling methods.\cite{ribeiro2018achieving} In addition, in recent work that is soon to be published we have applied RAVE to study the kinetics of ligand dissociation.\cite{yihang_info_bottleneck}

\section{CASE STUDIES}

\subsection{Sub-microsecond unbinders}
\citet{caflisch_plos} have studied the dissociation of six small ligands from the FKBP protein using unbiased MD simulations, which was possible due to their choice of ligands whose dissociation was expected to occur in $\sim$20 ns,\cite{caflisch_plos} based on experimental dissociation constants\cite{burkhard2000x}. It was found that the kinetics for these dissociations followed single-exponential kinetics and that the unbinding times ranged between four and eighteen ns.\cite{caflisch_plos} An interesting result from their work is the observation that the ligands can take various conformations at different FKBP binding site positions and that there exists a heterogeneity of dissociation paths. In addition, \citet{pan2017jctc} performed long unbiased MD simulations that were $\mu$s in length with the Anton supercomputer on the same protein (FKBP) with various small ligands but with a different force-field.\cite{pan2017jctc} Several spontaneous association and dissociation events were observed and $k_{on}$ and $k_{off}$ estimated. Similar to the conclusion of \citet{caflisch_plos}, ligand binding to different binding sites were observed, and multiple dissociation paths were sampled.\cite{pan2017jctc} The reported  residence times in 8--140 ns range\cite{pan2017jctc} which interestingly is quite different from the numbers reported by some of these ligands in Ref. \cite{caflisch_plos}. 

\subsection{Benzamidine-trypsin} 
The benzamidine-trypsin system has been one of the most popular systems for simulating the protein-ligand association/dissociation process using all-atom MD simulations, starting with the MSM based work in 2011.\cite{trypsin_msm} Since then numerous other methods have been applied in the last few years to study this system, with the only benchmark\cite{guillain1970use} that has been used in these studies being a 1970 experimental measurement of $k_{off}$=600 $s^{-1}$. MSM seem to be systematically overestimating the experimental rate constant, with predicted $k_{off}$ of 95000 $s^{-1}$ in 2011\cite{trypsin_msm}, 28000 $s^{-1}$ in 2014\cite{doerr2014fly} and more recently 13000 $s^{-1}$ in 2015\cite{plattner2015protein}. Use of WE method gives 5555 $s^{-1}$, AMS gives 240  $s^{-1}$ while metadynamics gives 9 $s^{-1}$. Thus we have different MD methods giving rates that range from 160 times faster than experiments to 65 times slower than experiments -- or a 4 orders of magnitude variation. This is a strikingly large range, likely arising from a combination of the use of different force-fields and different sampling methods. What is however more re-assuring that many of these different methods using different force-fields have obtained similar dominant unbinding pathway of this system, with similar metastable states as defined by the ligand position, orientation, protein conformation, and water molecule location. This more pronounced agreement between pathways than between rate constants is worthy of further discussion and is revisited in Sec. \ref{discussion}. 

\subsection{Protein kinases}
Kinases are arguably the second most important group of drug targets, after G-protein-coupled receptors. Thus naturally there has been a strong interest in applying all-atom MD methods to study the association/dissociation of pharmaceutical ligands from protein kinases. In 2011, the special-purpose supercomputer Anton was used to study the association dynamics of the drug Dasatinib to Src kinase\cite{shaw_dasatanib} where they were able to achieve spontaneous binding of the ligand (but not unbinding) in around 1 out of 6 trajectories. They found numerous attractor spots on the surface of the protein, and the binding process essentially involved navigating the maze of these short-lived albeit metastable states on the way to the actual bound pose. They also identified intriguing behavior of water molecules during binding, where a single layer of water molecules (lifetime 0.1 $\mu$s) was found to be one of the many impediments to binding. Later, Tiwary et al performed infrequent metadynamics simulation of the unbinding of this same system, where they achieved full unbinding in 12 out of 12 simulations, identifying metastable states and their fluxes.\cite{dasat_koff} They also found waters to play a very critical role. Namely, they found that a well-preserved Glu-Lys salt bridge had to distort before the ligand could exit the protein, aided by solvent water molecules. As we describe in Sec. \ref{discussion}, this finding has important repercussions for possible use of constant pH methods in the simulation of drug unbinding. The calculated association and dissociation rate constants from the Anton and metadynamics simulations respectively were in good agreement with experiments, and perhaps more significantly, with each other. In a similar vein, Casasnovas et al have applied metadynamics to study unbinding of urea-based BIRB family inhibitor from p38 kinase.\cite{p38} In this last work, they performed metadynamics using two different sets of CVs, and using the self-consistent measure of Ref. \cite{pvalue}, they were able to select one of these as reliable. The $k_{off}$ value from this agreed well with experiments, demonstrating the usefulness of the test from Ref. \cite{pvalue} in ascertaining the reliability of dynamics reconstructed from metadynamics.

\subsection{Heat shock protein 90}
Heat shock protein 90 (HSP90) is a chaperone protein generally found in all bacterial and eukaryotic cells, and inhibitors of HSP90 have strong potential to be used as anti-cancer drugs. However, the presence of various conformational states and flexibility of the HSP90 protein at the binding site makes the investigation of kinetics very challenging\cite{bruce2018review}. Mollica \textit{et al.} used the smoothed-potential MD or scaled MD which is essentially an approximate and computationally efficient version of infrequent metadynamics, with the intent to focus more on relative dissociation rates rather than absolute rates. They used this to study dissociation of several ligands from HSP90 and obtained agreement with reported experimental results. In a somewhat similar vein, Random acceleration molecular dynamics (RAMD) method has also been used to study binding/unbinding pathways of various proteins\cite{nowak2015jcp,tang2017pcm} and calculate relative residence times for a set of various ligands to the N-terminal domain of the HSP90, finding good agreement with reported results.

\subsection{G protein--coupled receptors (GPCRs)}
G protein--coupled receptors (GPCRs) are of tremendous pharmacological importance, with $\sim$50\% of all drugs on the market designed to act through modulation of GPCRs\cite{garland2013gpcr}. GPCRs are trans-membrane proteins comprised of seven $\alpha$-helices and have the critical cellular function of controlling the cell’s response to the presence of extracellular molecules. Due to both their fundamental and practical importance it is desirable to investigate the dynamics of GPCRs using all-atom MD. Using unbiased MD on the Anton supercomputer, Dror et al. studied the binding of antagonists and agonists to the $\beta_{2}$ adrenergic receptor,\cite{dror2011gpcr} observing twelve binding events leading to a $k_{on}$ estimate of 3.1$\times$10$^7$ M$^{-1}$s$^{-1}$, which is in good agreement with the experimental value of 1.0$\times$10$^7$ M$^{-1}$ s$^{-1}$. Notice however that the absolute $k_{off}$ has not been calculated for GPCR-ligand complexes, although \citet{mollica2015gpcr} used the dissociation process in order to rank four different triazine-based ligands according to their residence times, relative to a reference. Good agreement with experimental results was reported.\cite{mollica2015gpcr} In addition, \citet{bortolato2015gpcr} determined whether a series of ligands were fast or slow using their recent technique named adiabatic--biased metadynamics (aMetaD) in conjunction with an approximate residence time score. In more recent work, \citet{meral2018gpcr} used Metadynamics with a Maximum Caliber approach and reported that thermodynamic properties as well as kinetic properties can be achieved for GPCR activation.


\subsection{HIV-1 protease}
The HIV-1 protease contains two $\beta$-hairpin loops called flaps whose dynamics, according to different studies,\cite{karthik2011protein,huang2017biochemistry} play an important role in the mechanisms of ligand association and dissociation from the binding pockets. For this reason the HIV-1 protease is a challenging -- and interesting -- system for MD simulations. Focusing on investigating the mechanism of binding for a peptide with HIV-1 protease, Pietrucci \textit{et al.}\cite{pietrucci2009jacs} employed the bias-exchange metadynamics (BEMD) method and captured important features of the mechanism such as water bridges, conformational changes of the flaps, various important interactions between the ligand and protease. They calculated the kinetics from a 7-dimensional order parameter space through a weighted histogram method and the $k_{on}$ and $k_{off}$ values were in rough agreement with the available experimental results. Sun \textit{et al.}\cite{sun2017jcim} studied kinetics for 6 systems including HIV protease complexed with the drug lopinavir using infrequent metadynamics, finding decent overall agreement in rate constants.

\section{CONCLUSIONS: ARE WE THERE YET?}
\label{discussion}
In this short review we have described some of the methods useful for calculating the kinetics, and more generally the mechanisms, of ligand-protein dissociations from all-atom MD simulations, and highlighted some of their successes and shortcomings with case-studies. A clear message which is embedded throughout is that the use of MD simulations to investigate ligand-protein dissociations is brimming with potential and could soon have a significant impact in the rational design of drugs and in understanding biochemical processes fundamental to life. This is because MD simulations complement experimental kinetics in a natural way, since they provide a wealth of experimentally difficult-to-get mechanistic information. This being said, there are large obstacles remaining for unleashing the full potential of MD simulations to practical applications. The first obstacle is the inherently limited timescales that can be attained in all-atom MD simulations. We have highlighted here a number of ingenious MD methods that have been devised to deal with this limitation. This sampling problem is coupled to identifying a good RC, which we would like to argue is a limitation in the biased but also the unbiased methods, even thought it might not be obvious for the latter class of methods. Recent years have shown signs of hope in this RC problem through the development of many automated methods.\cite{ticameta,metatica,sgoop,rave} A second problem is the force-field issue. Due to the large number of atoms involved in biochemical systems, an analytic model, called force-field, describing the energy interactions between the constituents must be used. These models often are based on available structural and thermodynamic data, which tends to be richer in describing longer lived metastable structures. Keep in mind, however, that kinetic estimates depend on sampling a whole ensemble of transition paths which necessarily venture outside these data-rich regions. The careful cataloging of the transition states describing rare event processes should find use in helping refine the available force-fields, which would lead to better estimates from MD simulations since the predictions stemming from MD simulations depend on the accuracy of the force-field used to model the microscopic interactions between these atoms. In a sense, these two limitations are coupled -- developing force-fields that better characterize not just thermodynamic but also kinetic data essentially needs development of reliable sampling methods that can deal with long timescales.

In fact, one crucial metric that is missing thus far is the degree to which the errors in sampling are decoupled from the force-field errors. The benzamidine--trypsin case study serves as a perfect example, with different methods using different force-fields producing dissociation timescales differing by a staggering four orders of magnitude, even though somewhat reassuringly, many of these studies have discovered similar dissociation pathways and crucial intermediate states. Both could be significant in deviations of kinetic predictions from experimental values. A thorough, systematic investigation of how the same MD methods handle the different available force-fields for the same systems would be invaluable in providing a sound estimate of the errors from force-fields relative to the sampling. This will help in establishing how valid it is to compare $k_{off}$ and $k_{on}$ values directly with experiments. An important example of this is the different water models used in force-fields, which can lead to water diffusivities that are three to four times as different.\cite{tiwary2015pnas} Similarly, yet another parameter which has not received much mention in typical MD based publications for calculating dissociation timescales and mechanisms is the friction coefficient used in the simulation in order to maintain desired temperature. This coefficient can also have profound effects on the calculations through what is known as the Kramers' turnover phenomenon.\cite{thirumalai_kramers,kramers} Such differences can significantly affect the calculated rate constants and even lead to agreement between experiments and simulations for wrong reasons. 
Finally, we would like to highlight that some of the slowly dissociating systems have also been found through all--atom simulations to show salt--bridge distortions and thus possible protontation state dependence.\cite{shan2008conserved,dasat_koff,wallace2011continuous} This hints on the usefulness of constant pH simulations in the study of dissociation mechanisms. To conclude, all-atom MD simulations go beyond giving just a rate constant -- they hold the promise of giving pathways and unambiguously highlighting the role of individual players such as specific residues and individual water molecules. Much progress has been made in this field, but a long way remains before all-atom MD simulations can be deemed truly complimentary and even predictive of actual experiments of ligand--protein dissociation processes.

\textbf{Acknowledgments}
We thank Zachary Smith and Freddy Alexis Cisneros for their careful reading of this manuscript. We also thank Deepthought2, MARCC and XSEDE (projects CHE180007P and CHE180027P) for helping our group with computational resources used to develop some of the methods reported in this work. Pratyush Tiwary would like to thank the University of Maryland Graduate School for financial support through the Research and Scholarship Award (RASA). Yihang Wang would like to thank the NCI-UMD partnership for Integrative Cancer Research for financial support. 
\bibliography{tiwary_references}

\end{document}